\documentclass[11pt,english]{article}

\usepackage{geometry}
\usepackage{setspace}
\usepackage{amssymb}
\usepackage{custom_tex}
\title{Proximal Causal Inference without Uniqueness Assumptions}
\begin{document}

%\begin{frontmatter}


\author{Jeffrey Zhang, Wei Li, Wang Miao, Eric Tchetgen Tchetgen}
\maketitle


%\author[1]{Jeffrey Zhang}
%\author[2]{Wei Li}
%\author[3]{Wang Miao}
%\author[1]{Eric Tchetgen Tchetgen}

%\affiliation[1]{organization={Department of Statistics and Data Science, The Wharton School, The University of Pennsylvania},%Department and Organization state={PA},country={U.S.A.}}
        
%\affiliation[2]{organization={Center for Applied Statistics and School of Statistics, Renmin University of China}, country={P.R.C.}}
%\affiliation[3]{organization={Department of Probability and Statistics,Peking University}, country={P.R.C.}}

\begin{abstract}
We consider identification and inference about a counterfactual outcome mean when there is unmeasured confounding using tools from proximal causal inference (\cite{miao}, \cite{intro_proxy}). Proximal causal inference requires existence of solutions to at least one of two integral equations. We motivate the existence of solutions to the integral equations from proximal causal inference by demonstrating that, assuming  the existence of a solution to one of the integral equations, $\sqrt{n}$-estimability of a linear functional (such as its mean) of that solution requires the existence of a solution to the other integral equation. Solutions
to the integral equations may not be unique, which complicates estimation and inference. We construct a consistent estimator for the solution set for one of the integral equations and then adapt
the theory of extremum estimators to find from the estimated set a consistent estimator
for a uniquely defined solution. A debiased estimator for the counterfactual mean is shown to be root-$n$ consistent, regular, and asymptotically semiparametrically locally efficient  under additional regularity conditions.
\end{abstract}



%\begin{keyword}
%Proximal Causal Inference \sep $\sqrt{n}$-estimability



%\end{keyword}

%\end{frontmatter}


\newtheorem{assumption}{Assumption}
\section{Introduction}
It is widely acknowledged that unmeasured confounding is pervasive in observational studies, as it is unlikely that an investigator will have measured all confounders of the treatment and outcome. Often, the most one can hope for is that some measured confounders can act as proxies for true, unmeasured confounders. Proximal causal inference was developed to circumvent the issue of unmeasured confounders through the use of suitable proxy variables. See \cite{miao}, \cite{intro_proxy}, \cite{selective_review}, and \cite{shi_jrssb} for a more comprehensive overview of the framework. Proximal causal inference leverages the existence of either a treatment confounding bridge function or an outcome confounding bridge function, which solve certain integral equations (\cite{miao}, \cite{semi_proxy}, \cite{deaner}, \cite{kallus}).  Then, the population average treatment effect (ATE) and the average treatment effect on the treated (ATT) are respectively uniquely identified nonparametrically as a certain linear functional of a confounding bridge function, without the latter necessarily being uniquely identified (\cite{miao}, \cite{semi_proxy}, \cite{deaner}, \cite{kallus}). However, construction of a root-$n$ consistent, regular and asymptotically linear semiparametric locally efficient estimator of the ATE or ATT in prior literature has relied exclusively on uniqueness of the bridge functions with a notable exception being the concurrent work of \cite{bennett2022}. In this note, the goal is to investigate estimation and inference of the counterfactual mean, and thus of the ATE and ATT, when uniqueness does not hold. Specifically, we construct a root-$n$ consistent, regular and asymptotically linear nonparametric estimator of the ATE without requiring uniqueness. The proposed methods build on recent de-biasing methods developed in \cite{SANTOS2011129} and \cite{nonig}. In somewhat related settings, the former considers a linear functional of the structural function in (i) the widely studied nonparametric instrumental variable problem (\cite{chen_2012}), while the latter considers (ii) a nonparametric shadow variable framework in a nonignorable missing data problem (\cite{dhaul}, \cite{miao_shadow}).
Proximal identification and inference differs from both of these settings in that the identification challenge requires the use of two proxies, while (i) and (ii) technically require a single proxy, mainly a valid instrument in (i) and a valid shadow variable in (ii). De-biasing methods have also been utilized for other purposes in causal inference, see for example \cite{auto_debias}.

An outline of the paper is as follows:
in Section \ref{identifcation}, we review identification strategies for the counterfactual mean, and thus the ATE and ATT, from previous works under the proximal framework, and describe sufficient and necessary conditions for identification and root-$n$ estimation of the ATE. In Section \ref{estimation_strat}, we develop an estimator for the counterfactual mean, and therefore for the ATE and ATT, establish the asymptotic theory,  and discuss its semiparametric efficiency.  We conclude
with a discussion in Section \ref{discussion} and include proofs in the Supplementary Material. While for ease of exposition, all results are given for the counterfactual outcome mean, they equally apply to a broader class of functionals as discussed in the Supplementary Material. 
\section{Identification}
\label{identifcation}
We wish to estimate the effect of a binary treatment $A$ on an outcome $Y$ in a setting with unmeasured confounding. Define $Y(a)$ for $a=0,1$ to be the potential outcomes of the response if treatment had been externally set to $a$. The goal is to estimate the average treatment effect, $\E[Y(1)-Y(0)]$. Let $U$ be an unmeasured confounder and $X$ a vector of observed covariates. Statistical independence is denoted by $\perp$. Instead of assuming no unmeasured confounding, we adopt the recent proximal causal inference framework wherein we require there to be a treatment confounding proxy $Z$ and an outcome confounding proxy $W$. This leads to the following assumptions as introduced by \cite{semi_proxy}:
\begin{assumption}
\label{cons}
(Consistency) $Y=AY(1)+(1-A)Y(0)$, almost surely.
\end{assumption}
\begin{assumption}
\label{lat_unc}
(Latent Unconfoundedness)  $(Z,A) \perp (Y(a),W) | U,X \text{ for } a = 0, 1.$ 
\end{assumption}
\begin{assumption}
\label{pos2}
(Positivity) $0 < \mathbb{P}(A = a|U, X) < 1$ almost surely, for $a = 0, 1$.
\end{assumption}
\begin{assumption}
\label{comp1}
(Completeness 1)
\begin{enumerate}
    \item For any $a, x$, if $\E[g(U)|Z, A = a, X = x] = 0$ almost surely, then $g(U) = 0$ almost
surely.
    \item For any $a, x$, if $\E[g(Z)|W, A = a, X = x] = 0$ almost surely, then $g(Z) = 0$ almost surely.
\end{enumerate}
\end{assumption}
There are two ways to identify the counterfactual mean $\E[Y(a)]$. The first method is as follows:
\begin{theorem}
\label{miao}
(\cite{miao}) Suppose that there exists an outcome confounding bridge
function $h(w, a, x)$ that solves the following integral equation
\begin{equation}
\label{eq:1}
    \E[Y |Z, A, X] = \int
h(w, A, X)dF(w|Z, A, X),
\end{equation}
almost surely. Then, under Assumptions \ref{cons}, \ref{lat_unc},\ref{pos2}, and \ref{comp1}(1), one has that
\begin{comment}
\begin{equation}
\label{eq:2}
\E[Y |U, A, X] = \int
h(w, A, X)dF(w|U, X)
\end{equation}
almost surely. Moreover, the counterfactual mean $\E[Y (a)]$ is nonparametrically identified by
\end{comment}
\begin{equation}
\label{eq:3}
\E[Y (a)] = \int_{\mathcal{X}}  \int h(w, a, x)dF(w|x)dF(x).
\end{equation}
\begin{comment}
and thus the average treatment effect is identified by
\begin{equation}
\label{eq:4}
\psi = \int_{\mathcal{X}}  \int
[h(w, 1, x) - h(w, 0, x)]dF(w|x)dF(x)
\end{equation}
\end{comment}
\end{theorem}

\begin{assumption}
\label{comp2}
(Completeness 2) 
\begin{enumerate}
    \item 
For any a, x, if $\E[g(U)|W, A = a, X = x] = 0$ almost surely, then $g(U) = 0 $ almost
surely.
\item For any a, x, if $\E[g(W)|Z, A = a, X = x] = 0$ almost surely, then $g(W) = 0$ almost
surely.
\end{enumerate}
\end{assumption}
Using the second completeness assumption provides an alternative identification scheme:
\begin{theorem}
\label{cui}
(\cite{semi_proxy}) Suppose that there exists a treatment confounding bridge function $q(z, a, x)$
that solves the integral equation
\begin{equation}
\label{eq:5}
    \E[q(Z, a, X)|W, A = a, X] = \frac{1}{\mathbb{P}(A = a|W, X)}.
\end{equation}
Then, under Assumptions \ref{cons}, \ref{lat_unc}, \ref{pos2}, and \ref{comp2}(1), one has
\begin{comment}
\begin{equation}
\label{eq:6}
    \int q(z, a, X)dF(z|U, A = a, X) = \frac{1}{\mathbb{P}(A = a|U, X)}
\end{equation}
almost surely. Moreover, $\E[Y (a)]$ is nonparametrically identified by
    
\end{comment}
\begin{equation}
\label{eq:7}
\E[Y (a)] = \int_{\mathcal{X}} \int
I(\tilde{a} = a)q(z, a, x)ydF(y, z, \tilde{a}|x)dF(x).
\end{equation}
\begin{comment}
Therefore, the average treatment effect is identified by
\begin{equation}
\label{eq:8}
    \psi = \int_{\mathcal{X}}  \int
(-1)^{1-a}
q(z, a, x)ydF(y, z, a|x)dF(x)
\end{equation}
\end{comment}
\end{theorem}
Only existence of an outcome confounding bridge function $h$ or treatment confounding bridge function $q$ is required for identification of the ATE; they need not be unique. Suppose that one has observed $n$ i.i.d. data samples consisting of variables  $(A,Z,X,W,Y)$. Let $L_2(Y,W,Z,A,X)$ denote the space of real valued functions of $(Y,W,Z,A,X)$ that are square integrable with respect to the distribution of $(Y,W,Z,A,X)$ and use the inner product $ \langle f_1,f_2\rangle:=\E[f_1f_2]$. For any bounded linear map $T$, define $\mathcal{D}(T), \mathcal{R}(T), \mathcal{N}(T), T'$ to be the domain, range, null space, and adjoint of $T$. Let $T^{\perp}$ be the orthocomplement of a set $T$. Let the suffix $o$ denote relationship to the outcome confounding bridge function and let the suffix $tr$ denote relationship to the treatment confounding bridge function. Accordingly, define $T_o:L_2(W,A,X) \to L_2(Z,A,X)$ where $T_o g(W,A,X)=\E[g(W,A,X)|Z,A,X]$. Let $T_{tr}:L_2(Z,A,X) \to L_2(W,A,X)$ where $T_{tr} g(Z,A,X)=\E[g(Z,A,X)|W,A,X]$.
Before proceeding with inference, we provide a purely statistical motivation for the integral equations \ref{eq:1} and \ref{eq:5}. The following requires no reference to an unmeasured confounder $U$ or the causal structure.
Consider the following two scenarios:
\begin{enumerate}
    \item Suppose (\ref{eq:1}) holds, i.e. there exists a function $h \in L_2(W,A,X)$ such that $\E[Y|Z,A,X]=\E[h(W,A,X)|Z,A,X]$.
    \item Suppose (\ref{eq:5}) holds, i.e. there exists a function $q \in L_2(Z,A,X)$ such that $\frac{1}{\mathbb{P}(A|W,X)}=\E[q(Z,A,X)|W,A,X]$.
\end{enumerate}
Under the first scenario, consider the problem of estimating a functional of the following form:
\begin{equation}
\label{wfunctional}
    \beta_o=\E[\phi_o(W,A,X)h(W,A,X)],
\end{equation}
where $\phi_o$ is a known function in $L_2(W,A,X)$.
  Then we have the following:
\begin{proposition}
\label{id1}
Under the assumption that \ref{eq:1} holds, 
$\beta_o$ is identified iff $\phi_o \in \mathcal{N}(T_o)^{\perp}$. 
\end{proposition}
\begin{proof}
First, suppose $\beta_o$ is identified. Consider $h_1$, $h_2$ that satisfy Equation $\ref{eq:1}$. Note that this implies that $h_1-h_2 \in \mathcal{N}(T_o)$. Since $\beta_o$ is identified, both $h_1$ and $h_2$ yield the same value of $\beta_o$. Thus, we have that 
$0=\E[\phi_o(W,A,X)(h_1-h_2)]$
and so $\phi_o(W,A,X) \in \mathcal{N}(T_o)^{\perp}$ since $h_1-h_2$ is an arbitrary element of $\mathcal{N}(T_o)$. Conversely, suppose $\phi_o(W,A,X) \in \mathcal{N}(T_o)^{\perp}$. Then for any $h_1$ and $h_2$ that satisfy Equation \ref{eq:1}, we have $h_1-h_2 \in \mathcal{N}(T_o)$ and so $\E[\phi_o(W,A,X)h_1]=\E[\phi_o(W,A,X)h_2]$ and so $\beta_o$ is identified.
\end{proof}
Note that $\mathcal{N}(T_o)^{\perp}=\text{cl}(\mathcal{R}(T_o'))$. However, the following Lemma establishes that $\phi_o(W,X) \in \mathcal{R}(T_o')$ is necessary for $\beta_o$ to be $\sqrt{n}$ estimable.
\begin{lemma}
\label{rootnh}
Assuming equation \ref{eq:1} holds and regularity condition \ref{regularityh} in the appendix, $\phi_o(W,X) \in \mathcal{R}(T_o')$ is necessary for $\beta_o$ to be $\sqrt{n}$ estimable.
\end{lemma}
This result is analogous to a result derived in \cite{SEVERINI2012491} in the non-parametric instrumental variables context.
Next, note that
\begin{align*}
    \E[h(W,a,X)]  &= \E[\E[h(W,a,X)|W,X]] \\ &= \E[\E[h(W,a,X)I(A=a)/\mathbb{P}(A=a|W,X)|W,X]] \\  &= \E[h(W,A,X)/\mathbb{P}(A=a|W,X)I(A=a)]
\end{align*}
which is in the form of equation \ref{wfunctional} with  $\phi_o(W,A,X)=I(A=a)/\mathbb{P}(A=a|W,X)$ which for current purposes  may be assumed known.  Lemma \ref{rootnh} thus implies that for $\E[h(W,a,X)]$ to be $\sqrt{n}$ estimable, there must be a function $q(Z,A,X)$ that satisfies
\begin{equation*}
    \E[q(Z,A,X)|W,A,X]=I(A=a)/\mathbb{P}(A=a|W,X).
\end{equation*}
This corresponds to the condition from Equation \ref{eq:5}. 
Likewise, consider the problem of estimating a functional of the following form:
\begin{equation}
\label{zfunctional}
\beta_{tr}=\E[\phi_{tr}(Z,A,X)q(Z,A,X)],
\end{equation}
where $\phi_{tr}$ is a known function in $L_2(Z,A,X)$. Analogous to Proposition \ref{id1}, we have the following:
\begin{proposition}
\label{id2}
Under the assumption that \ref{eq:5} holds, $\beta_{tr}$ is identified iff $\phi_{tr} \in \mathcal{N} (T_{tr})^{\perp}$.
\end{proposition}
\begin{proof}
Note that for any $q_1$ and $q_2$ that satisfy \ref{eq:5}, we must have $q_1-q_2 \in \mathcal{N}(T_{tr})$. Then the argument follows in the exact same manner as in Proposition \ref{id1}.
\end{proof}
As above, it is possible to establish that $\phi_{tr} \in \mathcal{R}(T_{tr}')$ is necessary for $\beta_{tr}$ to be $\sqrt{n}$ estimable.
\begin{lemma}
\label{rootnq}
Assuming equation \ref{eq:5} holds and regularity condition \ref{regularityq} in the appendix, $\phi_o(W,X) \in \mathcal{R}(T_o')$ is necessary for $\beta_o$ to be $\sqrt{n}$ estimable.
\end{lemma}
Next, observe that from Equation \ref{eq:7}, we have 
\begin{align*}
    \E[I(A=a)q(Z,a,X)Y]  &= \E[\E[I(A=a)q(Z,a,X)Y|Z,A,X]] \\ &= \E[I(A=a)q(Z,a,X)\E[Y|Z,A,X]]
    \\ &= \E[I(A=a)q(Z,A,X)\E[Y|Z,A,X]]
\end{align*}
which is in the form of equation \ref{zfunctional} with $\phi_{tr}(Z,A,X)=I(A=a)\E[Y|Z,A,X]$ which for current purposes  may be assumed known.  Lemma \ref{rootnq} thus implies that for $\E[I(A=a)q(Z,A,X)Y]$ to be $\sqrt{n}$ estimable, there must be a function $h(W,A,X)$ such that
\begin{equation*}
    \E[h(W,A,X)|Z,A,X]=I(A=a)\E[Y|Z,A,X].
\end{equation*}
This corresponds to the condition from Equation \ref{eq:1}. We may conclude that taking as a primitive condition that a solution to Equation \ref{eq:1} exists everywhere in the model, i.e. at all laws included in the semiparametric model, identification and root-n estimation of the counterfactual outcome mean necessarily implies that a solution to \ref{eq:5} exists at the true data generating law. On the other hand, taking as a primitive condition that a solution to Equation \ref{eq:5} exists at all laws of the semiparametric model, identification and root-n estimation of the counterfactual outcome mean necessarily implies that a solution to \ref{eq:1} exists at the true data generating law. 

The present setting differs from the shadow variable missing data setting studied in \cite{nonig} in ways worth discussing. In the current setting, we aim to account for the presence of an unmeasured confounder $U$ and the key assumption \ref{lat_unc} to identification involves this latent variable together with two fully observed auxiliary factors in the form of a pair of proxies $Z$ and $W$, each of which plays a specific role. In contrast, identification in a shadow variable setting does not require invoking a latent factor, and  requires only a single fully observed auxiliary variable which satisfies a certain conditional independence condition (\cite{nonig}). Despite these differences, our paper demonstrates that the analytic framework of \cite{nonig} readily extends to the proximal causal inference setting. We further establish in the Supplementary Material that the approach actually applies to a general class of doubly robust functionals studied by \cite{ghassami}, for which a pair of nuisance functions are defined as solutions to Fredholm integral equations. The above propositions give formal justification that certain Fredholm integral equations must admit a solution in order for a regular estimator of functionals in this class to exist. In the next section, we describe an estimation strategy for the counterfactual mean without the assumption of a unique $h$ or $q$ that solve the integral equations. 

\section{Estimation Strategy}
\label{estimation_strat}
We follow estimation strategies from \cite{SANTOS2011129} and \cite{nonig}. Based on the above discussion, it is sensible to construct solution sets for either of the Fredholm integral equations from equation \ref{eq:1} and \ref{eq:5}. Without loss of generality, we consider estimating the solution set of equation \ref{eq:1}. First, let $\mathcal{H}$ be a set of smooth functions. Define the solution sets of the Fredholm integral equations as follows:
\begin{equation}
    \mathcal{H}_0 = \{h \in \mathcal{H} : \E[h(W,A,X)|Z,A,X]=\E[Y|Z,A,X]\}.
\end{equation}
Under the assumptions from Theorem \ref{miao}, $\E[h(W,a,X)]$ has a causal interpretation as the counterfactual mean $\E[Y(a)]$. Under these assumptions, to estimate $\mu_a:=\E[Y(a)]$, we first construct a consistent estimator $\wh{\mathcal{H}}_0$ for the set $\mathcal{H}_0$; next, we choose a specific $\wh{h}_0 \in \wh{\mathcal{H}}_0$ so that it is a consistent estimator for a fixed element $h_0 \in \mathcal{H}_0$. 
\subsection{Estimation of solution sets}
Define the criterion function
\begin{align*}
    C(h)&=\E[\{\E[Y-h(W,A,X)|Z,A,X]^2\}].
\end{align*}
In practice, the estimation of the solution set can be done in the two arms separately, for example, by taking the criterion function $C(h)=\E[I(A=a)\E[(Y-h(W,a,X))|Z,A=a,X]^2] $. Note that 
$\mathcal{H}_0=\{h \in \mathcal{H} : C(h)=0\},$
i.e., the solution set of the Fredholm integral equation consists of the zeros of the criterion function. To proceed with estimation, we adopt a two-stage approach. We aim to construct sample analogues $C_n$ of the criterion function $C$. We let $\mathcal{H}_n$ be sieve for $\hh$. Specifically, for a known sequence of approximating functions $\{\psi_m(w,a,x)\}_{m=1}^\infty$, let 
\begin{equation}
\begin{aligned}
    \mathcal{H}_n&=\left\{h \in \hh : h(w,a,x) = \sum_{m=1}^{m_n}\beta_m \psi_m(w,a,x)\right\},
\end{aligned}
\end{equation}
for $\beta,h$ unknown and $m_n$ known.
To construct $C_n$, we require a nonparametric estimator of conditional expectations. For this, let $\{\phi_k(z,a,x)\}_{k=1}^\infty$ be a known sequence of approximating functions. Denote
\begin{equation*}
\begin{aligned}
    \mathbf{\phi}(z,a,x)=\{\phi_1(z,a,x),\ldots,\phi_{k_n}(z,a,x)\}^T, \\
    \mathbf{\Phi}=\{\mathbf{\phi}(Z_1,A_1,X_1),\ldots,\mathbf{\phi}(Z_n,A_n,X_n)\}^T.
\end{aligned}
\end{equation*}

For a generic random variable $B=B(W,A,X,Y,)$ with realizations $\{B_i=B(W_i,A_i,X_i,Y_i)_{i=1}^n\}$ the nonparametric sieve estimator of $\E[B|a,z,x]$ is 
\begin{equation}
    \wh{\E}(B|A,Z,X)=\mathbf{\phi}(Z,A,X)(\mathbf{\Phi}^T\mathbf{\Phi})^{-1}\sum_{i=1}^n \mathbf{\phi(Z_i,A_i,X_i)}B_i.
\end{equation}
The sample analogue $C_n(h)$ is then
\begin{equation}
    C_n(h)=\frac{1}{n}\sum_{i=1}^n\wh{e}^2(Z_i,A_i,X_i,h),
\text{ where }
    \wh{e}(Z_i,A_i,X_i,h)=\wh{\E}[Y-h(W,A,X)|A_i,Z_i,X_i].
\end{equation}
Then the proposed estimator of $\hh_0$ is $\wh{\hh}_0=\{h \in \hh_n: C_n(h)\leq c_n \}$
where $c_n$ is an appropriately chosen sequence that tends to 0. 
\subsection{Set consistency}
In this section, we establish the set consistency of $\wh{\mathcal{H}}_0$ for $\mathcal{H}_0$ under Hausdorff distances. For this, define the Hausdorff distance between two sets $\hh_1,\hh_2 \subset \hh$ as 
\begin{equation*}
    d_H(\hh_1,\hh_2, ||\cdot||)=\text{max}\{d(\hh_1,\hh_2),d(\hh_2,\hh_1)\},
\end{equation*}
where $d(\hh_1,\hh_2)=\text{sup}_{h_1 \in \hh_1}\inf_{h_2 \in \hh_2}||h_1-h_2||$ and $||\cdot|| $ is a given norm. Consider the following two norms:
\begin{align*}
    ||h||_w^2&=\E[\{\E[h(W,A,X)|Z,A,X]\}^2], \ ||h||_\infty^2=\sup_{w,a,x}|h(w,a,x)|.
\end{align*}
We say that a set of estimators $\widehat{K}$ is said to be \emph{consistent} for a set $K$ if $d_H(\widehat{K}, K,\|\cdot\|) \rightarrow_p 0.$
Next, notice that any $h,h_0 \in \hh_0$ satisfy \ref{eq:1}, so it holds that for any $\wh{h}_0 \in \wh{\hh}_0$ and any $h, h_0 \in \hh_0$, $||\wh{h}_0-h_0||_w=||\wh{h}_0-h||_w$ and so 
\begin{equation}
    ||\wh{h}_0-h_0||_w=\inf_{h \in \hh_0}||\wh{h}_0-h||_w \leq d_H(\wh{\hh}_0,\hh_0,||\cdot||_w).
\end{equation}
Thus, we can calculate the convergence rate of $||\wh{h}_0-h_0||_w$ by finding the convergence rate of $d_H(\wh{\hh}_0,\hh_0,||\cdot||_w)$. We will need to consistently estimate $\hh_0$ under the supremum norm because under the $||\cdot||_w$ norm, elements of $\hh_0$ form an equivalence class. We will require several assumptions.
\begin{assumption}
\label{compact} 
%(Analogous to Assumption 3 from \cite{nonig})
The vector of covariates $X \in \RR^d$ has support $[0,1]^d$, and outcome $Y \in \mathbb{R}$ and proxies $Z,W \in \RR$ have compact support.
\end{assumption}
\begin{definition}
For a generic function $\rho(\omega)$ defined on $\omega \in \R^d$ we define
\begin{equation*}
    \|\rho\|_{\infty, \alpha}=\max _{|\lambda| \leq \underline{\alpha}} \sup _{w}\left|D^{\lambda} \rho(w)\right|+\max _{\lambda=\underline{\alpha}} \sup _{w \neq w^{\prime}} \frac{D^{\lambda} \rho(w)-D^{\lambda} \rho\left(w^{\prime}\right)}{\left\|w-w^{\prime}\right\|^{\alpha-\underline{\alpha}}},
\end{equation*}
where $\lambda$ is a $d$-dimensional vector of nonegative integers, $|\lambda|=\sum_{i=1}^d \lambda_i$, $\underline{\alpha}$ denotes the
largest integer smaller than $\alpha$, $D^{\lambda} \rho(\omega)=\partial^{|\lambda|}\rho(\omega)/\partial \omega_1^{\lambda_1}\ldots \partial \omega_d^{\lambda_d}$, and $D^{0}\rho(\omega)=\rho(\omega)$.
\end{definition}
\begin{assumption}
\label{H_approx}
%(Analogous to Assumption 4 from \cite{nonig})
The following conditions hold:
\begin{enumerate}[(i)]
    \item $\sup_{h \in \hh} ||h||_{\infty,\alpha} < \infty$ for some $\alpha > (d+1)/2$; $\hh_0 \neq \emptyset$, $\hh_n$ and $\hh$ are closed.
    \item for every $h \in \hh$, there is a $\Pi_n h \in \hh_n$ such that $\sup_{h \in \hh}||h-\Pi_n h||_{\infty} = O(\eta_n)$ for some $\eta_n=o(1)$.
\end{enumerate}
\end{assumption}
\begin{assumption}
\label{Phi_approx}
%(Analogous to Assumption 5 from \cite{nonig})
The following conditions hold:
\begin{enumerate}[(i)]
\item The smallest and largest eigenvalues of $\E[\phi(Z,A,X)\phi(Z,A,X)^T]$ are bounded above and away from zero for all $k_n$.
\item for every $h \in \hh$, there is a $\mathbf{\pi}_n(h) \in \RR^{k_n}$ such that \\$\sup_{h\in\hh} ||\E[h(W,A,X)|a,z,x]-\phi^T(z,a,x)\mathbf{\pi}_n(h)||_{\infty}=O\left(k_n^{-\frac{\alpha}{d+1}}\right).$
\item $\xi_n^2k_n=o(n)$, where $\xi_n=\sup_{z,a,x}||\phi(z,a,x)||_2$.
\end{enumerate}
\end{assumption}
Then we have the following proposition:
\begin{proposition}
\label{prop:2.1}
%(Analogous to Theorem 3.2 from \cite{nonig})
Suppose that Assumptions \ref{compact}-\ref{Phi_approx} hold. If $a_n=O\lambda_n^{-1}$, $b_n \to \infty$, and $b_n=o(a_n)$, then
$$d_H(\wh{\hh}_0,\hh_0,\Norm{\cdot}_\infty)=o_p(1) \text{ and } d_H(\wh{\hh}_0,\hh_0,\Norm{\cdot}_w)=O_p(c_n^{1/2}).$$
\end{proposition}
After obtaining a consistent estimator $\wh{\hh}_0$ of $\hh_0$, we select a specific estimator from $\wh{\hh}_0$ such that it converges to a unique element in $\hh_0$. 
\subsection{A representer-based estimator}
To do so, we let $M:\hh\to \R$ be a population criterion function that attains a unique minimum $h_0$ on $\hh_0$ and $M_n(h)$ its sample analogue. We then select
\begin{equation}
\label{eq:15}
    \wh{h}_0\in \argmin_{h \in \wh{\hh}_0}M_n(h).
\end{equation}
We make the following assumption about $M$:
\begin{assumption}
\label{M}
%(Analogous to Assumption 6 from \cite{nonig})
The function set $\hh$ is convex; the functional $M:\hh \to \R$ is strictly convex and attains a unique minimum at $h_0$ on $\hh_0$; its sample analogue $M_n: \hh \to \R$ is continuous and $\sup_{h\in \hh}|M_n(h)-M(h)|=o_p(1)$. 
\end{assumption}
A sensible choice for $M$ is the squared length $M(h)=\E[h(W,A,X)^2]$
 with corresponding sample analog $M_n(h)=\frac{1}{n}\sum_{i=1}^n h(W_i,A_i,X_i)^2$.

\begin{proposition}
\label{prop:2.2}
%(Analogous to Theorem 3.3 from \cite{nonig})
Suppose that assumptions \ref{compact}-\ref{M} hold. Then $\Norm{\wh{h}_0-h_0}_\infty=o_p(1),$
where $\wh{h}_0$ is defined in equation \ref{eq:15}. If $a_n=O(\lambda_n^{-1})$, $b_n\to \infty$ and $b_n=o(a_n)$, we then have 
$\Norm{\wh{h}_0-h_0}_w=O_p(C_n^{1/2}).$
\end{proposition}
Based on the identification $\mu_a=\E[Y(a)]=\E[h_0(W,a,X)]$, we propose the following estimator for $\mu_a$:
\begin{equation}
    \wh{\mu}_a=\frac{1}{n}\sum_{i=1}^n\cb{\wh{h}_0(W_i,a,X_i)}.
\end{equation}
To facilitate analysis of the asymptotic properties of the estimator, let $\overline{\hh}$ be the closure of the linear span of $\hh$ under $\Norm{\cdot}_w$, and define
\begin{equation*}
    \la h_1,h_2 \ra_w=\E\sqb{I(A=a)\E\{h_1(W,A,X)|A,Z,X\}\E\{h_2(W,A,X)|A,Z,X\}}.
\end{equation*}
Next, we require the following:
\begin{assumption}
\label{rep}
%(Analogous to Assumption 7 from \cite{nonig})
The following conditions hold:
\begin{enumerate}[(i)]
    \item there exists a function $g_0 \in \hh$ such that
    $\la g_0, h \ra_w=\E\{h(W,a,X)\},$ for all $h \in \overline{\hh}$.
    \item $\eta_n=o(n^{-1/3}), k_n^{-3\alpha/(d+1)}=o(n^{-1}), k_n^3=o(n), \xi_n^2k_n^2=o(n)$, and $\xi_n^2k_n^{-2\alpha/(d+1)}=o(1)$.
\end{enumerate}
\end{assumption}
Observe that any $g_0$ that satisfies Assumption \ref{rep}(i) satisfies the following for all $h \in \overline{\hh}$:
\begin{equation}
\label{eq:ass_rep_imp}
    \E\sqb{I(A=a)\E\{g_0(W,A,X)|A,Z,X\}h(W,A,X)} = \E[I(A=a)/\mathbb{P}(A=a\mid W,X)h(W,A,X)].
\end{equation}
Then suppose that $\E\sqb{I(A=a)\E\{g_0(W,A,X)|A,Z,X\}-I(A=a)/\mathbb{P}(A=a\mid W,X)\mid W,A,X} \in \overline{\hh}.$ Using \ref{eq:ass_rep_imp}, this implies that 
\begin{equation}
\label{repisaq}
    \E[I(A=a)(\{g_0(W,A,X)|A,Z,X\}-1/\mathbb{P}(A=a\mid W,X))\mid W,A,X]=0.
\end{equation}
In other words, $\E[\{g_0(W,A,X)|A,Z,X\}]$ solves the integral equation from \ref{eq:5}. Thus, Assumption \ref{rep}(i) can be viewed as a strengthening of the assumption that there exists a function $q(Z,A,X)$ that solves Equation \ref{eq:5}.
As in \cite{SANTOS2011129}, the $g_0$ function will be unique up to equivalence class. In the Supplementary Material, we present a theorem characterizing the asymptotic expansion of $\wh{\mu}_a$, which includes the non-negligible term $r_n(\wh{h}_0)$ defined below: 
\begin{equation}
r_n(\wh{h}_0)=\frac{1}{n}\sum_{i=1}^n  I(A_i=a)\wh{\E}\{\Pi_n g_0(W,A,X)|A_i,Z_i,X_i\}\wh{e}(Z_i,A_i,X_i,\wh{h}_0).
\end{equation}

\begin{comment}
Observe that a sufficient condition for Assumption \ref{rep}(i) to hold is 
\begin{equation}
\label{eq:17}
    \E\{g_0(W,A,X)|A,Z,X\}=I(A=a)q(Z,A,X)
\end{equation}
for \emph{some} $q(Z,A,X)$ that satisfies the equation $\E[q(Z,A,X)|W,A,X]=\frac{1}{P(A=a|W,X)}$. To see this, note that if \ref{eq:17} holds,
\begin{equation*}
\begin{aligned}
    \la g_0, h\ra_w&= \E\sqb{\E\{g_0(W,A,X)|A,Z,X\}\E\{h(W,A,X)|A,Z,X\}} \\ &=\E\sqb{I(A=a)q(Z,A,X)\E\{h(W,A,X)|A,Z,X\}} \\ &= \E\sqb{\E\{I(A=a)q(Z,A,X)h(W,A,X)|A,Z,X\}} \\ &=\E[I(A=a)q(Z,A,X)h(W,A,X)] \\ &= \E\sqb{\E[I(A=a)q(Z,A,X)h(W,A,X)|A,W,X]} \\ &= \E[I(A=a)/P(A=a|W,X)h(W,A,X)]\\ &= \E[h(W,a,X)]
\end{aligned}
\end{equation*}
\end{comment}

\begin{comment}
\begin{theorem}
\label{thm:2.3}
%(Analogous to Theorem 3.4 from \cite{nonig})
Suppose that assumptions \ref{compact}-\ref{rep} hold. Then we have that 
\begin{equation*}
\begin{aligned}
    \sqrt{n}(\wh{\mu}_a-\mu_a)&=\frac{1}{\sqrt{n}}\sum_{i=1}^n\sqb{h_0(W_i,a,X_i)-\mu_a+I(A_i=a)\E\{g_0(W,A,X)|A_i,Z_i,X_i\}\times(Y_i-h_0(W_i,A_i,X_i))}\\&-\sqrt{n}r_n(\wh{h}_0)+o_p(1),
    \end{aligned}
\end{equation*}
where 
\begin{equation}
r_n(\wh{h}_0)=\frac{1}{n}\sum_{i=1}^n  I(A_i=a)\wh{\E}\{\Pi_n g_0(W,A,X)|A_i,Z_i,X_i\}\wh{e}(Z_i,A_i,X_i,\wh{h}_0).
\end{equation}
\end{theorem}  
\end{comment}

\subsection{A de-biased estimator}
To get an asymptotically normal estimator, we will de-bias the estimator $\wh{\mu}_a$, which requires estimation of $r_n(\wh{h}_0)$. First, we define a new criterion function $R(h)$ and its sample analog $R_n(h)$:
\begin{equation*}
\begin{aligned}
    &R(h)=\E\sqb{I(A=a)\E\{h(W,A,X)|Z,A,X\}^2}-2\E\{h(W,a,X)\} \ , \forall h \in \hh, \\
    &R_n(h)=\frac{1}{n}\sum_{i=1}^nI(A_i=a)\wh{\E}\{h(W,A,X)|Z_i,A_i,X_i\}^2-\frac{2}{n}\sum_{i=1}^n h(W_i,a,X_i) \ , \forall h \in \hh.
\end{aligned}
\end{equation*}
Since $\la g_0, h \ra_w=\E\{h(W,a,X)\}$, we have that $R(h)=\Norm{h-g_0}^2_w-\Norm{g_0}_w^2$. It follows that $g_0$ is the unique minimizer of the mapping $h \to R(h)$. Then since $g_0$ is close to $\Pi_n g_0$ by assumption \ref{H_approx}(ii), we can estimate the term $\Pi_n g_0$ by 
\begin{equation}
    \wh{g} \in \argmin_{h \in \hh_n } R_n(h).
\end{equation}
With this estimate, we can construct the following estimator for $r_n(\wh{h}_0)$:
\begin{equation}
    \wh{r}_n(\wh{h}_0)=\frac{1}{n}\sum_{i=1}^n  I(A_i=a)\wh{\E}\{\wh{g}(W,A,X)|A_i,Z_i,X_i\}\wh{e}(Z_i,A_i,X_i,\wh{h}_0).
\end{equation}
Next, we have a lemma characterizing the convergence of $\wh{r}_n(\wh{h}_0)$ to $r_n(\wh{h}_0)$.
\begin{lemma}
\label{lem:2.4}
%(Analogous to Lemma 3.5 from \cite{nonig})
Suppose that assumptions \ref{compact}-\ref{Phi_approx} and \ref{rep} hold. Then 
\begin{equation*}
    \sup_{\wh{h}_0\in \wh{\hh}_0} \left|\widehat{r}_{n}\left(\widehat{h}_{0}\right)-r_{n}\left(\widehat{h}_{0}\right)\right|=O_{p}\left[c_{n}^{1 / 2}\left\{\left(\frac{k_{n}}{n}\right)^{1 / 4}+k_{n}^{-\frac{\alpha}{2(d+1)}}\right\}\right].
\end{equation*}
\end{lemma}
Using this lemma, we can construct the de-biased estimator $\wh{\mu}_{a-db}=\wh{\mu}_{a}+\wh{r}_n(\wh{h}_0)$.

\begin{theorem}
\label{thm:2.5}
%(Analogous to Theorem 3.6 in \cite{nonig})
Suppose assumptions \ref{compact}-\ref{rep} hold. If $a_n = O(\lambda_n^{-1})$, $b_n \to \infty$ and
$n^{2/3}b_n = o(a_n)$, then $\sqrt{n}(\wh{\mu}_{a-db} -\mu_a)$ converges in distribution to $N(0, \sigma^2)$, where $\sigma^2$
is
the variance of 
\begin{equation}
\label{eq:22}
h_0(W,a,X)-\mu_a+I(A=a)\E\{g_0(W,A,X)|Z,A,X\}\times(Y-h_0(W,A,X)).
\end{equation}
\begin{comment}
Moreover, a consistent estimator for the variance $\sigma^2$ is given by 
\begin{equation*}
\begin{aligned}
    &\frac{1}{n}\sum_{i=1}^n \sqb{ \wh{h}_0^2(W_i,a,X_i)-\cb{\frac{1}{n}\sum_{i=1}^n\wh{h}_0(W_i,a,X_i)}^2+ I(A_i=a)\{\wh{\E}(\wh{g}(W,X)|Z_i,A=a,X_i\}^2\{Y_i-\widehat{h}_0(W_i,a,X_i)\}^2}\\&+2
\end{aligned}
\end{equation*}
\end{comment}
\end{theorem}
Equation \ref{eq:22} is the influence function for the de-biased estimator $\wh{\mu}_{a-db}$.
 We present an immediate corollary of \ref{thm:2.5} regarding local efficiency of this estimator in the Supplementary Material. 
\section{Discussion}
Using the proximal causal inference framework, we have established an estimator for the counterfactual mean under unmeasured confounding. We have shown it is consistent and presented conditions for when it is asymptotically normal.  Note that if interest is in the average causal effect, $\E[Y(1)-Y(0)]$, the proposed methodology can be adapted by slightly adjusting \ref{rep}(i). 
\label{discussion}

\section*{Acknowledgements}
We thank Andrea Rotnitzky for pointing out some errors in the previous version.
J. Zhang was supported by NIH grant 5R01HD101415-02. W. Li's research was supported by the National Natural Science Foundation of China (NSFC 12101607), the Fundamental Research Funds for the Central Universities
and the Research Funds of Renmin University of China. E. Tchetgen Tchetgen (PI) was supported by NIH Grants: R01AI27271, R01CA222147, R01AG065276, R01GM139926.
\appendix

\section{Regularity conditions}
\begin{assumption}
\label{regularityh}
Let $h^*(W,A,X)=P_{\mathcal{N}(T_o)^{\perp}}h(W,A,X)$.
\begin{enumerate}[(i)]
    \item Denote $\Psi_o(z,a,x)=\E[(Y-h^*(W,A,X))^2|Z=z,A=a,X=x]$. Suppose $0 < \inf_{z,a,x} \Psi_o(z,a,x) < \sup_{z,a,x} \Psi_o(z,a,x) < \infty$.
    \item $T_o'$ is compact.
\end{enumerate}

\end{assumption}
\begin{assumption}
\label{regularityq} Let $q^*(Z,A,X)=P_{\mathcal{N}(T_{tr})^{\perp}}q(Z,A,X)$.
\begin{enumerate}[(i)]
    \item Denote $\Psi_{tr}(w,a,x)=\E[(q^*(Z,A,X))^2|W=w,A=a,X=x]$. Suppose $0 < \inf_{w,a,x} \Psi_{tr}(w,a,x) < \sup_{w,a,x} \Psi_{tr}(w,a,x) < \infty$.
    \item $T_{tr}'$ is compact.
\end{enumerate}

\end{assumption}

\bibliographystyle{plainnat}
\bibliography{template_short}

%\begin{thebibliography}{00}



%\end{thebibliography}
\end{document}

\maketitle
\appendix
In this supplementary material document, we first restate all assumptions from the main manuscript. Section \ref{appendix} collects additional lemmas and proofs of results in the main manuscript. Section \ref{appendix_class} outlines how the approach in the main manuscript could be adjusted so that it can be applied to a larger class of functionals.

\begin{assumption}
\label{cons}
(Consistency) $Y=AY(1)+(1-A)Y(0)$ almost surely.
\end{assumption}
\begin{comment}
\begin{assumption}
\label{pos1}
(Positivity)  $0 < \mathbb{P}(A = a|L) < 1$  $a = 0, 1$ almost surely.
\end{assumption}
\begin{assumption}
\label{exch}
(Exchangeability) $Y (a) \perp A|L$ for $a = 0, 1$.
\end{assumption}
\end{comment}
\begin{assumption}
\label{lat_unc}
(Latent Unconfoundedness)  $$(Z,A) \perp (Y(a),W) | U,X \text{ for } a = 0, 1.$$ 
\end{assumption}
\begin{assumption}
\label{pos2}
(Positivity) $0 < \mathbb{P}(A = a|U, X) < 1$ almost surely, $a = 0, 1$.
\end{assumption}
\begin{assumption}
\label{comp1}
(Completeness 1)
\begin{enumerate}
    \item For any $a, x$, if $\E[g(U)|Z, A = a, X = x] = 0$ almost surely, then $g(U) = 0$ almost
surely.
    \item For any $a, x$, if $\E[g(Z)|W, A = a, X = x] = 0$ almost surely, then $g(Z) = 0$ almost surely.
\end{enumerate}
\end{assumption}
\begin{assumption}
\label{comp2}
(Completeness 2) 
\begin{enumerate}
    \item 
For any a, x, if $\E[g(U)|W, A = a, X = x] = 0$ almost surely, then $g(U) = 0 $ almost
surely.
\item For any a, x, if $\E[g(W)|Z, A = a, X = x] = 0$ almost surely, then $g(W) = 0$ almost
surely.
\end{enumerate}
\end{assumption}
\begin{assumption}
\label{compact} 
%(Analogous to Assumption 3 from \cite{nonig})
The vector of covariates $X \in \RR^d$ has support $[0,1]^d$, and outcome $Y \in \mathbb{R}$ and proxies $Z,W \in \RR$ have compact support.
\end{assumption}
\begin{assumption}
\label{H_approx}
%(Analogous to Assumption 4 from \cite{nonig})
The following conditions hold:
\begin{enumerate}[(i)]
    \item $\sup_{h \in \hh} ||h||_{\infty,\alpha} < \infty$ for some $\alpha > (d+1)/2$; $\hh_0 \neq \emptyset$, $\hh_n$ and $\hh$ are closed.
    \item for every $h \in \hh$, there is a $\Pi_n h \in \hh_n$ such that $\sup_{h \in \hh}||h-\Pi_n h||_{\infty} = O(\eta_n)$ for some $\eta_n=o(1)$.
\end{enumerate}
\end{assumption}
\begin{assumption}
\label{Phi_approx}
%(Analogous to Assumption 5 from \cite{nonig})
The following conditions hold:
\begin{enumerate}[(i)]
\item The smallest and largest eigenvalues of $\E[\phi(Z,A,X)\phi(Z,A,X)^T]$ are bounded above and away from zero for all $k_n$.
\item for every $h \in \hh$, there is a $\mathbf{\pi}_n(h) \in \RR^{k_n}$ such that $$\sup_{h\in\hh} ||\E[h(W,A,X)|a,z,x]-\phi^T(z,a,x)\mathbf{\pi}_n(h)||_{\infty}=O\left(k_n^{-\frac{\alpha}{d+1}}\right).$$
\item $\xi_n^2k_n=o(n)$, where $\xi_n=\sup_{z,a,x}||\phi(z,a,x)||_2.$
\end{enumerate}
\end{assumption}
\begin{assumption}
\label{M}
%(Analogous to Assumption 6 from \cite{nonig})
The function set $\hh$ is convex; the functional $M:\hh \to \R$ is strictly convex and attains a unique minimum at $h_0$ on $\hh_0$; its sample analogue $M_n: \hh \to \R$ is continuous and $\sup_{h\in \hh}|M_n(h)-M(h)|=o_p(1)$. 
\end{assumption}
\begin{assumption}
\label{rep}
%(Analogous to Assumption 7 from \cite{nonig})
The following conditions hold:
\begin{enumerate}[(i)]
    \item there exists a function $g_0 \in \hh$ such that
    $$\la g_0, h \ra_w=\E\{h(W,a,X)\}$$ for all $h \in \overline{\hh}$.
    \item $\eta_n=o(n^{-1/3}), k_n^{-3\alpha/(d+1)}=o(n^{-1}), k_n^3=o(n), \xi_n^2k_n^2=o(n)$, and $\xi_n^2k_n^{-2\alpha/(d+1)}=o(1)$.
\end{enumerate}
\end{assumption}

\section{Lemmas and Proofs}
\label{appendix}
\begin{lemma} (Theorem B.1 in \cite{SANTOS2011129})
\label{lem:a1}
Assume
\begin{enumerate}[(i)]
    \item $C(h) \geq 0$ for any $h \in \hh$ with $\hh$ compact in some norm $\Norm{\cdot}$.
    \item $\hh_n \subset \hh$ are closed and $\sup_{\hh}\inf_{\hh_n}\Norm{h-h_n}=O(c_{1n})$. 
    \item Uniformly on $\hh_n$, $C_n(h)\leq C_1C(h)+O_p(c_{2n})$ and $C(h)\leq C_1C_n(h)+O_p(c_{2n})$ with probability approaching one.
    \item $C(h)\leq C_2 \inf_{h_0\in \hh_0}\Norm{h-h_0}^{\kappa_1}$ for some $\kappa_1>0$. 
    \\ \\ Then for $a_n=(\max\{c_{1n}^{\kappa_1},c_{2n}\}^{-1})$ and $b_n \to \infty$ with $b_n=o(a_n)$, the set $\wh{\hh}_0$ with $c_n=b_n/a_n$ satisfies $d_H(\wh{\hh}_0,\hh_0,\Norm{\cdot})=o_p(1)$. If in addition,
    \item $C(h)\geq \inf_{h_0 \in \hh_0}C_2\Norm{h-h_0}^{\kappa_2}$ for some $\kappa_2 > 0$.\\ \\ then $d_H(\wh{\hh}_0,\hh_0,\Norm{\cdot})=O_p(\max\{(b_n/a_n)^{1/\max(\kappa_2,1)},c_{2n}\})$.
\end{enumerate}
\end{lemma}
\begin{lemma}
\label{lem:a2}
(Lemma A.2 in \cite{nonig})
For $\omega:[0,1]^d\times \yy \times \cw \to \RR$, let $\overline{\E}[\omega(X,Y,W)|A=a,z,x]=\phi^T(z,a,x)(\mathbf{\Phi}^T\mathbf{\Phi})^{-1}\sum_{i=1}^n\E\{\omega(X,A,Y,W)|A_i,Z_i,X_i \}$. If assumptions \ref{compact}-\ref{Phi_approx} hold, then
\begin{enumerate}[(i)]
    \item $\sup_{h \in \hh_n}\frac{1}{n}\sum_{i=1}^n\left[\wh{\E}\{h(W,A,X)|A_i,Z_i,X_i\}-\E\{h(W,A,X)|A_i,Z_i,X_i\}\right]^2=O_p\p{\frac{k_n}{n}+k_n^{-\frac{2\alpha}{d+1}}}.$
    \item $\sup_{h \in \hh_n} \E \sqb{\E\{Y-h(W,A,X)|A,Z,X\}-\overline{\E}\{Y-h(W,A,X)|A,Z,X\}}^2=O_p\p{k_n^{-\frac{2\alpha}{d+1}}}.$
\end{enumerate}
\end{lemma}
\begin{comment}
\begin{proof}
First, following \cite{NEWEY1997147}, under assumption \ref{Phi_approx}(i), we may assume without loss of generality that $\E\{I(A=a)\phi(Z,X)\phi(Z,X)^T\}=I$. Let $\gg_n$ be the linear span of $I(\tilde{a}=a)\phi(z,x)$. Then it follows that for any $g(\tilde{a},z,x)=I(\tilde{a}=a)\sum_{j=1}^{k_n}\beta_j\phi_j(z,x) \in \mathcal{G}_n$, we have
\begin{equation*}
    \frac{\Norm{g}_\infty}{\Norm{g}_{L^2}}=\sup_{\tilde{a},z,x}\frac{\left|I(\tilde{a}=a)\sum_{j=1}^{k_n}b_j\phi_j(z,x)\right|}{(\sum_{j=1}^{k_n}n_j^2)^{1/2}}\leq \sup_{z,x}\Norm{\phi(z,x)}
\end{equation*}
Let $A_n=\sup_{\gcal}\Norm{g}_\infty/\Norm{g}_{L^2}$. Then assumption \ref{Phi_approx}(ii) implies that $A_n^2k_n/n \to 0$. Following Huang, we have that with probability tending to one,
\begin{equation}
    \frac{1}{2}\E[g^2(A,Z,X)] \leq \frac{1}{n}\sum_{i=1}^n g^2(A_i,Z_i,X_i) \leq \E[g^2(A,Z,X)]
\end{equation}. 
\end{proof}
\end{comment}
\begin{corollary}
\label{cor:a3}
(Corollary A.1 in \cite{nonig})
If assumptions \ref{compact}-\ref{Phi_approx} hold, the uniformly in $\hh_n$, with probability tending to one, we have 
\begin{enumerate}[(i)]
    \item $C_n(h) \leq 8C(h)+O_p(\tau_n),$
    \item $C(h) \leq 8C_n(h)+O_p(\tau_n),$
\end{enumerate}
for $\tau_n=\frac{k_n}{n}+k_n^{-\frac{2\alpha}{d+1}}$.
\end{corollary}

\begin{proof}[Proof of Proposition 2.1]
The proof follows from verifying conditions in Lemma \ref{lem:a1}. Conditions (i) and (ii) hold with $\Norm{\cdot}=\Norm{\cdot}_\infty$ and $c_{1n}=\eta_n$ by assumption \ref{H_approx}. Condition (iii) holds with $c_{2n}=-\frac{k_n}{n}+k_n^{\frac{-2\alpha}{d+1}}$ and $C_1=8$ by \ref{cor:a3}. Also, note that
\begin{equation*}
\begin{aligned}
    C(h)&=\E[\E[Y-h(W,A,X)|Z,A,X]^2] \\ &= \E[\E[h_0(W,A,X)-h(W,A,X)|Z,A,X]^2] \\ &\leq \inf_{h \in \hh_0}\Norm{h_0-h}_\infty^2
\end{aligned}
\end{equation*}
So condition (iv) holds with $C_2=1$ and $\kappa_1=2$. Then by \ref{lem:a1}, the first part of the proposition holds. For the second, we can verify the conditions (i)-(iii) for the norm $\Norm{\cdot}_w$. Also, $C(h)=\Norm{h-h_0}_w^2$ so conditions (iv) and (v) from \ref{lem:a1} hold with  $C1 = C2 = 1$ and $\kappa_1 = \kappa_2 = 2$. Then by Lemmma \ref{lem:a1}, 
\begin{equation*}
    d_H(\wh{\hh}_0,\hh_0)=O_p(\max\{c_n^{1/2},\eta_n\}).
\end{equation*}
Since $c_n^{1/2}/\eta_n \to \infty$, we get 
\begin{equation*}
    d_H(\wh{\hh}_0,\hh_0)=O_p(c_n^{1/2}).
\end{equation*}
\end{proof}
\begin{lemma}
\label{lem:a4}
(Theorem 3.2 in \cite{SANTOS2011129})
Suppose that the following conditions hold: 
\begin{enumerate}[(i)]
    \item  $\hh_0 \subset \hh$ is closed with
$\hh$ compact and $M : \hh \to \RR$ has a unique minimum on $\hh_0$ at $h_0$.
\item $\wh{\hh}_0 \subset \hh$ satisfies $d_H(\wh{\hh}_0, \hh_0, \Norm{\cdot}) = o_p(1)$.
\item $M_n : \hh \to \RR$ and $M : \hh  \to \RR$ are continuous.
\item 
$\sup_{h \in \hh} |M_n(h) - M(h)| = o_p(1)$.
\end{enumerate}
If $\wh{h}_0 \in \argmin_{h\in\wh{\hh}_0}  M_n(h)$, then $\Norm{\wh{h}_0-h_0}= o_p(1).$
\end{lemma}
\begin{proof}[Proof of Proposition 2.2]
We show the first claim of the proposition by verifying conditions of Lemma \ref{lem:a4}. First, by Assumption \ref{H_approx}, both $\hh_0$ and $\hh$ are compact under $\Norm{\cdot}_\infty$, and hence condition (i) of Lemma \ref{lem:a4} holds. Second, the convexity of $\hh_0$ and strict convexity of
$M$ implies that $M$ has a unique minimum on $\hh_0$. With additional conditions in the proposition,
Lemma \ref{lem:a4} implies that
$\Norm{\wh{h}_0-h_0}_\infty = o_p(1)$.
For the second claim of the proposition, we note that
\begin{equation*}
    \Norm{\wh{h}_0-h_0}_w \leq d_H(\wh{\hh}_0,\hh_0),\Norm{\cdot}_w).
\end{equation*}
This, combined with Proposition 2.1, shows the second claim.
\end{proof}
\begin{lemma}
\label{lem:a5} (Lemma A.4 in \cite{nonig})
Suppose assumptions \ref{compact} and \ref{H_approx}(i) hold. Let $\A_n$ be the $\sigma$-filed generated by $\{A_i,X_i,Z_i\}$ and let $\{W_{in}\}_{i=1}^n$ be a triangle array of random variables that is measurable with respect to $\A_n$. If $n^{-1}\sum_{i=1}^n W_{in}^2=O_p(1)$ and $\Norm{\wh{h}_0-h_0}_\infty=o_p(1)$, then it follows that
\begin{equation*}
    \frac{1}{\sqrt{n}}\sum_{i=1}^n W_{in}\sqb{\wh{h}_0(W_i,A_i,X_i)-h_0(W_i,A_i,X_i)-\E \{\wh{h}_0(W,A,X)-h_0(W,A,X)|A_i,Z_i,X_i\}}=o_p(1).
\end{equation*}
\end{lemma}
\begin{lemma}
\label{lem:a6} (Lemma A.5 in \cite{nonig})
Suppose assumptions \ref{compact} and \ref{H_approx}(i) hold. Let $\A_n$ be the $\sigma$-filed generated by $\{A_i,X_i,Z_i\}$ and let $\{W_{in}\}_{i=1}^n$ and $\{\widetilde{W}_{in}\}_{i=1}^n$ be triangle arrays of random variables that are measurable with respect to $\A_n$ such that $n^{-1/2}\sum_i (W_{in}-\widetilde{W}_{in})^2=O_p(1)$. If $\wh{h}_0\in \hh_n$ satisfies $\Norm{\wh{h}_0-h_0}_\infty=o_p(1)$ and $\Norm{\wh{h}_0-h_0}_w=o_p(n^{-1/4})$, then it follows that 
\begin{equation*}
    \frac{1}{\sqrt{n}}\sum_{i=1}^nW_{in}\{Y_i-\wh{h}_0(W_iA_i,X_i)\}= \frac{1}{\sqrt{n}}\sum_{i=1}^n\widetilde{W}_{in}\{Y_i-\wh{h}_0(W_i,A_i,X_i)\}+o_p(1).
\end{equation*}
\end{lemma}
\begin{lemma}
\label{lem:a7} (Lemma A.6 in \cite{nonig})
Suppose assumptions \ref{compact}, \ref{H_approx}(i), \ref{Phi_approx}, \ref{rep}(ii) hold. Let the function class $\gcal$ satisfy: for all $g \in \gcal$, $\Norm{g}_\infty \leq K$ for some $K>0$, $\E[g(W,A,X)|Z,A,X]=0$ and $\int_0^\infty \sqrt{\log N(\epsilon, \gcal,\Norm{\cdot}_\infty} d\epsilon < \infty$. If $\overline{\E}[g(W,A,X)|a,z,x]=\phi^T(z,a,x)(\mathbf{\Phi}^T\mathbf{\Phi})^{-1}\sum_{i=1}^n\E\{g(W,A,X)|A_i,Z_i,X_i \}$ then it holds that 
\begin{equation*}
    \sup_{\gcal, \hh_{n}}\left|\frac{1}{\sqrt{n}} \sum_{i=1}^{n} \left[\E\left\{h(W,A,X) \mid A_i, Z_{i}, X_{i}\right\}-\bar{\E}\left\{h(W,A,X) \mid A_i, X_{i}, Z_{i}\right\}\right] g\left(W_{i},A_i, X_{i}\right)\right|=o_{p}(1).
\end{equation*}
\end{lemma}

\begin{lemma}
\label{lem:a8} (Lemma A.7 in \cite{nonig})
Suppose assumptions \ref{compact}-\ref{Phi_approx} and \ref{rep} hold. Let $u =\pm g_0$ and $u_n = \Pi_n u $. If $\wh{h}_0 \in \hh_n$ satisfies $\Norm{\wh{h}_0-h_0}_w=o_p(n^{-1/4})$ and $\Norm{\wh{h}_0-h_0}_\infty=o_p(1)$ then
\begin{enumerate}[(i)]
\item $$\frac{1}{\sqrt{n}}\sum_{i=1}^n I(A_i=a)\E\{u(W,A,X)|Z_i,A_i,X_i\} \sqb{\wh{e}(X_i,A_i,Z_i,\wh{h}_0)-(Y_i-\wh{h}_0(W_i,A_i,X_i))}=o_p(1);$$
    \item $$\sqrt{n}\la u, h_0-\wh{h}_0 \ra_w=\frac{1}{\sqrt{n}}\sum_{i=1}^n I(A_i=a) \E\{u(W,A,X)|Z_i,A_i,X_i\} \sqb{\wh{e}(X_i,A_i,Z_i,\wh{h}_0)-\wh{e}(X_i,A_i,Z_i,h_0)}+o_p(1).$$
\end{enumerate}
\end{lemma}
\begin{theorem}
\label{thm:2.3}
%(Analogous to Theorem 3.4 from \cite{nonig})
Suppose that assumptions \ref{compact}-\ref{rep} hold. Then we have that 
\begin{equation*}
\begin{aligned}
    \sqrt{n}(\wh{\mu}_a-\mu_a)&=\frac{1}{\sqrt{n}}\sum_{i=1}^n\sqb{h_0(W_i,a,X_i)-\mu_a+I(A_i=a)\E\{g_0(W,A,X)|A_i,Z_i,X_i\}\times(Y_i-h_0(W_i,A_i,X_i))}\\&-\sqrt{n}r_n(\wh{h}_0)+o_p(1),
    \end{aligned}
\end{equation*}
where 
\begin{equation}
r_n(\wh{h}_0)=\frac{1}{n}\sum_{i=1}^n  I(A_i=a)\wh{\E}\{\Pi_n g_0(W,A,X)|A_i,Z_i,X_i\}\wh{e}(Z_i,A_i,X_i,\wh{h}_0).
\end{equation}
\end{theorem}  
\begin{proof}[Proof of Theorem A.9]
By assumption \ref{H_approx}(i) and Theorem 2.7.1 from \cite{van1996weak}, we have that $\hh$ is a Donsker class. In addition, since $\Norm{\wh{h}_0-h_0}_\infty=o_p(1)$, we have that $\sup_{w,a,x}\mid \wh{h}_0(w,a,x)-h_0(w,a,x)\mid = o_p(1)$. As a result, we have 
\begin{equation}
    \frac{1}{\sqrt{n}}\sum_{i=1}^n \sqb{\{\wh{h}_0(W_i,a,X_i)-h_0(W_i,a,X_i)\}-\E\{\wh{h}_0(W,a,X)-h_0(W,a,X)\}}=o_p(1).
\end{equation}
Thus, 
\begin{equation}
    \sqrt{n}(\wh{\mu}_a-\mu_a)=\frac{1}{\sqrt{n}}\sum_{i=1}^n \{h_0(W_i,a,X_i)-\mu_a\}+\sqrt{n} \E[\wh{h}_0(W,a,X)-h_0(W,a,X)]+o_p(1).
\end{equation}
By definition of $g_0$, 
\begin{equation*}
    \sqrt{n} \E[\wh{h}_0(W,a,X)-h_0(W,a,X)]=\sqrt{n}\la g_0, \wh{h}_0-h_0\ra_w.
\end{equation*}
Applying Lemma \ref{lem:a8}, we derive that 
\begin{equation}
\begin{aligned}
    \sqrt{n} &\E[\wh{h}_0(W,a,X)-h_0(W,a,X)] \\ &= \frac{1}{\sqrt{n}}\sum_{i=1}^n I(A_i=a)\E\{g_0(W,A,X)|Z_i,A_i,X_i\} \sqb{\wh{e}(X_i,A_i,Z_i,h_0)-\wh{e}(X_i,A_i,Z_i,\wh{h}_0)}+o_p(1) \\ &= \frac{1}{\sqrt{n}}\sum_{i=1}^n I(A_i=a)\E\{g_0(W,A,X)|Z_i,A_i,X_i\} \sqb{Y_i-h_0(W_i,A_i,X_i)}\\ &-\frac{1}{\sqrt{n}}\sum_{i=1}^n I(A_i=a)\E\{g_0(W,A,X)|Z_i,A_i,X_i\} \sqb{\wh{e}(X_i,A_i,Z_i,\wh{h}_0)}+o_p(1)
    \end{aligned}
\end{equation}
By Cauchy-Schwarz,
\begin{equation}
\begin{aligned}
    &\frac{1}{\sqrt{n}}\sum_{i=1}^n \E\{g_0(W,A,X)-\Pi_ng_0(W,A,X)|Z_i,A_i,X_i\} \wh{e}(X_i,A_i,Z_i,\wh{h}_0) \\ &\leq \sqb{\frac{1}{\sqrt{n}}\sum_{i=1}^n \E\{g_0(W,A,X)-\Pi_ng_0(W,A,X)|Z_i,A_i,X_i\}^2}^{1/2} \\ &\times \cb{\frac{1}{\sqrt{n}}\sum_{i=1}^n \wh{e}(X_i,A_i,Z_i,\wh{h}_0)^2}^{1/2} \\ &\lesssim n^{1/4}\Norm{g_0-\Pi_n g_0}_\infty \times n^{1/4} \cb{C_n(\wh{h}_0)}^{1/2}
    \end{aligned}
\end{equation}
Then by Corollary A.3, Assumptions \ref{H_approx}(ii) and \ref{rep}(ii), 
\begin{equation}
    \frac{1}{\sqrt{n}}\sum_{i=1}^n  I(A_i=a)\E\{g_0(W,A,X)-\Pi_ng_0(W,A,X)|Z_i,A_i,X_i\} \wh{e}(X_i,A_i,Z_i,\wh{h}_0)=o_p(1).
\end{equation}
In a similar vein, using Cauchy Schwarz, Lemma \ref{lem:a2}, and Corollary \ref{cor:a3}, and assumption \ref{rep}(ii) we get
\begin{equation}
     \frac{1}{\sqrt{n}}\sum_{i=1}^n  I(A_i=a)\sqb{\E\{\Pi_ng_0(W,A,X)|Z_i,A_i,X_i\}-\wh{\E}\{\Pi_ng_0(W,A,X)|Z_i,A_i,X_i\}} \wh{e}(X_i,A_i,Z_i,\wh{h}_0)=o_p(1).
\end{equation}
Then by combining the above equations, we get the result
\begin{equation*}
\begin{aligned}
    \sqrt{n}(\wh{\mu}_a-\mu_a)&=\frac{1}{\sqrt{n}}\sum_{i=1}^n\sqb{h_0(W_i,a,X_i)-\mu_a+I(A_i=a)\E\{g_0(W,A,X)|A_i,Z_i,X_i\}\times(Y_i-h_0(W_i,A_i,X_i))}\\&-\sqrt{n}r_n(\wh{h}_0)+o_p(1),
    \end{aligned}
\end{equation*}
where 
\begin{equation*}
r_n(\wh{h}_0)=\frac{1}{n}\sum_{i=1}^n  I(A_i=a)\wh{\E}\{\Pi_n g_0(W,A,X)|A_i,Z_i,X_i\}\wh{e}(Z_i,A_i,X_i,\wh{h}_0).
\end{equation*}
\end{proof}

\begin{proof}[Proof of Lemma 3.3]
First, by assumption \ref{H_approx}(ii), we have that 
\begin{equation*}
    \Norm{\wh{g}-\Pi_ng_0}_w^2 \leq 2 \Norm{\wh{g}-g_0}_w^2+O(\eta_n^2).
\end{equation*}
Observe that $R(\wh{g})-R(\Pi_n g_0)=\Norm{\wh{g}-g_0}_w^2-\Norm{\Pi_n g_0-g_0}_w^2$. Thus we have
\begin{equation*}
    \begin{aligned}
\left\|\widehat{g}-\Pi_{n} g_{0}\right\|_{w}^{2} & \leq 2\left\{R(\widehat{g})-R\left(\Pi_{n} g_{0}\right)\right\}+O\left(\eta_{n}^{2}\right) \\
& \leq 2\left\{R(\widehat{g})-R_{n}(\widehat{g})+R_{n}(\widehat{g})-R_{n}\left(\Pi_{n} g_{0}\right)+R_{n}\left(\Pi_{n} g_{0}\right)-R\left(\Pi_{n} g_{0}\right)\right\}+O\left(\eta_{n}^{2}\right) \\
& \leq 4 \sup _{h \in \hh_{n}}\left|R(h)-R_{n}(h)\right|+O\left(\eta_{n}^{2}\right)
\end{aligned}
\end{equation*}
Next, we have 
\begin{equation}
    \begin{aligned}
    &\sup_{h \in \hh_n} \mid R(h)-R_n(h) \mid
    \\ &\lesssim \sup_{h \in \hh_n} \left| \E\sqb{\E\{h(W,A,X)|Z,A,X\}^2}-\frac{1}{n}\sum_{i=1}^n\wh{\E}\{h(W,A,X)|Z_i,A_i,X_i\}^2 \right| \\ &+ \sup_{h \in \hh_n}\left| \E[h(W,a,X)]-\frac{1}{n}h(W_i,a,X_i)\right| \\ &= \sup_{h \in \hh_n} \left| \E\sqb{\E\{h(W,A,X)|Z,A,X\}^2}-\frac{1}{n}\sum_{i=1}^n\wh{\E}\{h(W,A,X)|Z_i,A_i,X_i\}^2 \right| \\ &+ O_p(n^{-1/2})
    \end{aligned}
\end{equation}
since $\hh$ is a Donsker class as argued in the proof of Theorem 2.3. Next, define
\begin{equation*}
    \mathcal{F}=\{f(a,z,x)=I(\E[h(W,A,X)|a,z,x]^2:h \in \hh\}.
\end{equation*}
Note that we have for any $h_1,h_2 \in \hh$
\begin{equation*}
    \begin{aligned}
&\left|\left[E\left\{h_{1}(W, A,X) \mid a, x, z\right\}\right]^{2}-(\left[E\left\{h_{2}(W, A, X) \mid a, x, z\right\}\right]^{2}\right| \\
\leq &\left|E\left\{h_{1}(W,A, X)-h_{2}(W, A,X) \mid a, x, z\right\}\right| \times\left|E\left\{h_{1}(W, A,X)+h_{2}(W,A, X) \mid a, x, z\right\}\right| \\
\lesssim &\left\|h_{1}-h_{2}\right\|_{\infty}
\end{aligned}
\end{equation*}
Then by Theorems 2.7.11, 2.7.1, and 2.5.6 from \cite{van1996weak}, $\mathcal{F}$ is a Donsker class. So
\begin{equation}
    \sup_{h \in \hh_n} \left| \frac{1}{n}\sum_{i=1}^n\E\sqb{h(W,A,X)|Z_i,A_i,X_i}^2-\E\sqb{\E\sqb{h(W,A,X)|Z,A,X}^2}\right|=O_p(n^{-1/2}).
\end{equation}
We next bound the quantity:
\begin{equation*}
    \begin{aligned}
    &\sup_{h \in \hh_n} \left| \frac{1}{n}\sum_{i=1}^n \sqb{\E[h(W,A,X)|Z_i,A_i,X_i]}^2-\frac{1}{n}\sum_{i=1}^n \sqb{\wh{\E}[h(W,A,X)|Z_i,A_i,X_i]}^2 \right| \\ &\equiv \sup_{h \in \hh_n} \left| \frac{1}{n}(B_{i1}^2-B_{i2}^2)\right|
    \end{aligned}
\end{equation*}
We can bound 
\begin{equation*}
\begin{aligned}
\sup_{h \in \hh_n} \left| \frac{1}{n}\sum_{i=1}^n(B_{i1}^2-B_{i2}^2)\right| &\lesssim \sup_{h \in \hh_n}  \frac{1}{n}\sum_{i=1}^n(B_{i1}-B_{i2})^2+\sup_{h \in \hh_n} \left| \frac{1}{n}\sum_{i=1}^n B_{i1}(B_{i2}-B_{i1})\right| \\ &\lesssim \sup_{h \in \hh_n}  \frac{1}{n}\sum_{i=1}^n(B_{i1}-B_{i2})^2+\sup_{h \in \hh_n} \p{\frac{1}{n}\sum_{i=1}^nB_{i1}^2}^{1/2} \p{\frac{1}{n}\sum_{i=1}^n (B_{i2}-B_{i1})^2}^{1/2}
\end{aligned}
\end{equation*}
Then by assumption \ref{H_approx}(i) and Lemma \ref{lem:a2}, we have
\begin{equation*}
\begin{aligned}
    &\sup_{h \in \hh_n}\left|\frac{1}{n}\sum_{i=1}^n \sqb{\E[h(W,A,X)|A_i,Z_i,X_i]}^2-\frac{1}{n}\sum_{i=1}^n \sqb{\wh{\E}[h(W,A,X)|A_i,Z_i,X_i]}^2\right| \\ &=O_p\cb{(k_n/n)^{1/2}+k_n^{-\alpha/(d+1)}}
\end{aligned}
\end{equation*}
This equation with equations 29 and 30 imply that 
\begin{equation*}
    \sup_{h \in \hh_n} \left|R(h)-R_n(h) \right|=O_p\cb{(k_n/n)^{1/2}+k_n^{-\alpha/(d+1)}},
\end{equation*}
and so 
\begin{equation*}
    \sup_{h \in \hh_n} \Norm{\wh{g}-g_0}_w^2=O_p\cb{(k_n/n)^{1/2}+k_n^{-\alpha/(d+1)}}.
\end{equation*}
Finally, by Cauchy-Schwartz, we get 
\begin{equation*}
\begin{aligned}
    &\sup_{h \in \hh_n}  \sqrt{n} \mid \wh{r}_n(\wh{h}_0)-r_n(\wh{h}_0)\mid \\ & \leq \sqb{\frac{1}{n}\sum_{i=1}^n \{\wh{\E}(\wh{g}(W,A,X)-\Pi_n g_0(W,A,X)|A_i,Z_i,X_i)\}^2}^{1/2}\times\sup_{h \in \hh_n}\{C_n(\wh{h}_0\}^{1/2} \\ &\lesssim \sqb{\cb{\frac{1}{n}\sum_{i=1}^n\{\E(\wh{g}(W,A,X)-\Pi_n g_0(W,A,X)|A_i,Z_i,X_i)\}^2}^{1/2}+O_p\cb{(k_n/n)^{1/2}+k_n^{-\alpha/(d+1)}}}\\ & \times c_n^{1/2} \\ &\lesssim \sqb{\sup_{h \in \hh_n}\Norm{\wh{g}-\Pi_n g_0}_w+O_p\cb{(k_n/n)^{1/2}+k_n^{-\alpha/(d+1)}}}\times c_n^{1/2} \\ &= O_{p}\left[c_{n}^{1 / 2}\left\{\left(\frac{k_{n}}{n}\right)^{1 / 4}+k_{n}^{-\frac{\alpha}{2(d+1)}}\right\}\right]
\end{aligned}
\end{equation*}
where the second line follows from Lemma \ref{lem:a2}.
\end{proof}

\begin{proof}[Proof of Theorem 3.4]
Under assumption \ref{rep}, we have $\left(\frac{k_n}{n} \right)^{1/4}=o(n^{-1/6})$ and $k_n^{-\frac{\alpha}{2(d+1)}}=o(n^{-1/6})$. In addition, since $n^{2/3}b_n=o(a_n)$, by Lemma \ref{lem:a6}
\begin{equation*}
    \sup_{\wh{h}_0\in \wh{\hh}_0} \sqrt{n} \mid \wh{r}_n(\wh{h}_0)-r_n(\wh{h}_0)\mid =o_p(1).
\end{equation*}
This combined with Theorem A.9 implies the result.
\end{proof}

\begin{assumption}
\label{regularityh}
\
\begin{enumerate}[(i)]
    \item Denote $\Psi_o(z,a,x)=\E[(Y-h^*(W,A,X))^2|Z=z,A=a,X=x]$. Suppose $0 < \inf_{z,a,x} \Psi_o(z,a,x) < \sup_{z,a,x} \Psi_o(z,a,x) < \infty$.
    \item $T_o'$ is compact.
\end{enumerate}
where $h^*(W,A,X)=P_{\mathcal{N}(T_o)^{\perp}}h(W,A,X)$.
\end{assumption}
\begin{assumption}
\label{regularityq}
\
\begin{enumerate}[(i)]
    \item Denote $\Psi_{tr}(w,a,x)=\E[(q^*(Z,A,X))^2|W=w,A=a,X=x]$. Suppose $0 < \inf_{w,a,x} \Psi_{tr}(w,a,x) < \sup_{w,a,x} \Psi_{tr}(w,a,x) < \infty$.
    \item $T_{tr}'$ is compact.
\end{enumerate}
where  $q^*(Z,A,X)=P_{\mathcal{N}(T_{tr})^{\perp}}q(Z,A,X)$.
\end{assumption}
\begin{proof}[Proof of Lemma 2.4]
Let $t$ index a regular parametric submodel for the data which includes the true data-generating mechanism at $t=0$. We have that 
\begin{equation*}
    \E_t[h_t^*(W,A,X)|Z,A,X]=\E_t[Y|Z,A,X],
\end{equation*} for all t. \cite{semi_proxy} demonstrated that the tangent space assuming Equation 1 is as follows:
\begin{align*}
    \Lambda_1+\Lambda_2 &\equiv \{S(Z,A,X) \in L_2(Z,A,X): \E[S(Z,A,X)]=0\} \\ &+ \{S(Y,W|Z,A,X) \in L_2(Z,A,X)^{\perp}: \E[\epsilon S(Y,W|Z,A,X)|Z,A,X]\in \text{cl}(R(T_o)\},
\end{align*}
where $\epsilon=Y-h^*(W,A,X)$. They also showed that 
\begin{equation*}
    \E[\epsilon S(Y,W|Z,A,X)|Z,A,X]=\E[\partial h_t(W,A,X)/\partial t|_{t=0}|Z,A,X],
\end{equation*}
or in the notation of \cite{SEVERINI2012491}, 
\begin{equation*}
    \E[\epsilon S(Y,W|Z,A,X)|Z,A,X]=T_o\partial h_t(W,A,X)/\partial t|_{t=0},
\end{equation*}
which implies that 
\begin{equation*}
    T_o^+\E[\epsilon S(Y,W|Z,A,X)|Z,A,X]=\partial h^*_t(W,A,X)/\partial t|_{t=0},
\end{equation*}
where $ T_o^+$ is the Moore-Penrose pseudo-inverse of $T_o$. Consider the class of functionals $\beta_o=\E[\phi_o(W,A,X)h(W,A,X)]$. Following \cite{SEVERINI2012491}, it is possible to show that 
\begin{equation*}
    \partial \beta_o / \partial t |_{t=0}= \E[\phi_o \partial h_t(W,A,X)/\partial t|_{t=0} ]+\E[\phi_o h^*(W,A,X)S(Y,Z,A,W,X)].
\end{equation*}
Next, we follow the proof of Lemma 4.1 in \cite{SEVERINI2012491}.  Let $(\lambda_j,a_j,b_j)$ denote the singular system for $T_o$ and $T_o'$ where $(a_j)$ and $(b_j)$ are orthonormal bases for $\mathcal{N}(T_o)^{\perp}$ and $\text{cl}(\mathcal{R}(T_o))$, and $\lambda_j$ are singular values. We then have $\phi_o=\sum_{j=1}^\infty F_{a_j}(\phi_o)a_j$ where $F_{a_j}(\phi_o):=\la \phi_o,a_j\ra$. Then
\begin{equation*}
    \E[\phi_o T_o^+\E[\epsilon S(Y,W|Z,A,X)|Z,A,X] = \sum_{j=1}^\infty \lambda_j^{-1} F_{a_j}(\phi_o)F_{b_j}(\E[\epsilon S(Y,W|Z,A,X)|Z,A,X]) =: K_{\phi_o}(S).
\end{equation*}
Following arguments in \cite{SEVERINI2012491}, $K_{\phi_o}(S)$ is well-defined on $\Lambda_2$. Note that any $\phi_o$ that makes $K_{\phi_o}(S)$ unbounded will make $\partial \beta_o / \partial t |_{t=0}$ unbounded, and so the corresponding functional $\E[\phi_o h^*(W,A,X)]$ will not be $\sqrt{n}$-estimable. Let $S \in \Lambda_2$. Then by Cauchy-Schwarz and Bessel's inequality, 
\begin{equation*}
\begin{aligned}
|K_{\phi_o}(S)|^2 &\leq \sum_{j=1}^\infty \lambda_j^{-2}F_{a_j}^2(\phi_o)\Norm{\E[\epsilon S(Y,W|Z,A,X)|Z,A,X]}_{L_2(Y,W,Z,A,X)}^2 \\ &\lesssim \sum_{j=1}^\infty \lambda_j^{-2}F_{a_j}^2(\phi_o)\Norm{ S(Y,W|Z,A,X)}_{L_2(Y,W,Z,A,X)}^2
\end{aligned}
\end{equation*}
Since $\mathcal{R}(T_o')= \{S \in L_2(W,A,X):\sum_{j=1}^\infty \lambda_j^{-2}\la S, a_j \ra^2 < \infty\}$ by SVD of $T_o'$, $K_{\phi_o}(S)$ is bounded whenever $\phi_o \in \mathcal{R}(T_o')$. Next let, $\phi_o^0 \in \text{cl}(\mathcal{R}(T_o')) - \mathcal{R}(T_o')$. Then $\sum_{j=1}^\infty \lambda_j^{-2}F_{a_j}^2(\phi_o^0) = \infty$. By assumption \ref{regularityh}, for any $r$,
\begin{equation*}
    d_r:=(Y-h^*(W,A,X)) \Psi_o \sum_{i=1}^r \lambda_i^{-1}F_{a_i}(\psi_0)b_i \in L_2(Y,W,Z,A,X).
\end{equation*}
Moreover, $\E[Y-h^*(W,A,X)) \Psi_o \sum_{i=1}^r \lambda_i^{-1}F_{a_i}(\psi_0)b_i|Z,A,X]=0 $ and $\E[(Y-h^*(W,A,X))^2 \Psi_o \sum_{i=1}^r \lambda_i^{-1}F_{a_i}(\psi_0)b_i|Z,A,X] \in \text{cl}(\mathcal{R}(T_o)) $. Thus, $d_r \in \Lambda_2$. Also, we see that $K_{\phi_o}(d_r)=\Norm{d_r}_{L_2(Y,W,Z,A,X)}^2$. Then letting $f_r=d_r/\Norm{d_r}$, we see that $f_r$ is a sequence of unit vectors in $\Lambda_2$ such that \begin{equation*}
    \lim_{r\to \infty} K_{\phi_o^0}(f_r)= \left(\sum_{j=1}^\infty \lambda_j^{-2}F_{a_j}^2(\phi_o^0)\right)^{1/2} = \infty,
\end{equation*}
and so $K_{\phi_o^0}(S)$ is unbounded.
\end{proof}

\begin{proof}[Proof of Lemma 2.6]
Let $t$ index a regular parametric submodel for the data which includes the true data-generating mechanism at $t=0$. We have that 
\begin{equation*}
    \E_t[q_t^*(W,A,X)|Z,A,X]=1/P_t(A=a|W,X).
\end{equation*} for all t. First, note that following \cite{kallus}, the tangent space assuming Equation 3 is as follows: 
\begin{align*}
    \Lambda_1+\Lambda_2 &\equiv \{S(W,X) \in L_2(W,X): \E[S(W,X)]=0\} \\ &+ \{S(Z,A|W,X) \in L_2(W,X)^{\perp}: \E[q^*(Z,A,X) S(Z,A|W,X)|W,X]\in \text{cl}(R(T_{tr})\}.
\end{align*} Further, they showed that 
\begin{equation*}
    \E[q^*(Z,A,X)S(Z,A|W,X)|W,A,X]=-\E[\partial q_t(Z,A,X)/\partial t|_{t=0}|W,A,X],
\end{equation*}
or in the notation of \cite{SEVERINI2012491}, 
\begin{equation*}
    \E[q^*(Z,A,X)S(Z,A|W,X)|W,A,X]=-T_{tr}\partial q_t(Z,A,X)/\partial t|_{t=0},
\end{equation*}
which implies that 
\begin{equation*}
    T_{tr}^+\E[q^*(Z,A,X)S(Z,A|W,X)|W,A,X]=-\partial q_t(Z,A,X)/\partial t|_{t=0},
\end{equation*}
where $T_{tr}^+$ is the Moore-Penrose pseudo-inverse of $T_{tr}$. Consider the class of functionals $\beta_{tr}=\E[\phi_{tr}(Z,A,X)q(Z,A,X)]$. Following \cite{SEVERINI2012491}, it is possible to show that 
\begin{equation*}
    \partial \beta_{tr} / \partial t |_{t=0}= \E[\phi_{tr} \partial q_t(W,A,X)/\partial t|_{t=0} ]+\E[\phi_{tr} q^*(Z,A,X)S(Z,A,W,X)].
\end{equation*}
Next, we follow the proof of Lemma 4.1 in \cite{SEVERINI2012491}. Let $(\lambda_j,a_j,b_j)$ denote the singular system for $T_{tr}$ and $T_{tr}'$ where $(a_j)$ and $(b_j)$ are orthonormal bases for $\mathcal{N}(T_{tr})^{\perp}$ and $\text{cl}(\mathcal{R}(T_{tr}))$, and $\lambda_j$ are singular values. We then have $\phi_{tr}=\sum_{j=1}^\infty F_{a_j}(\phi_{tr})a_j$ where $F_{a_j}(\phi_{tr}):=\la \phi_{tr},a_j\ra$. Then
\begin{equation*}
    \E[\phi_{tr} T_{tr}^+\E[q^*(Z,A,X) S(Z,A|W,X)|W,A,X] = \sum_{j=1}^\infty \lambda_j^{-1} F_{a_j}(\phi_o)F_{b_j}(\E[q^*(Z,A,X) S(Z,A|W,X)|W,A,X]) =: K_{\phi_{tr}}(S).
\end{equation*}
Following arguments in \cite{SEVERINI2012491}, $K_{\phi_{tr}}(S)$ is well-defined on $\Lambda_2$.
Note that any $\phi_{tr}$ that makes $K_{\phi_{tr}}(S)$ unbounded will make $\partial \beta_{tr} / \partial t |_{t=0}$ unbounded, and so the corresponding functional $\E[\phi_{tr} q^*(Z,A,X)]$ will not be $\sqrt{n}$-estimable. Let $S \in \Lambda_2$. Then by Cauchy-Schwarz and Bessel's inequality, 
\begin{equation*}
\begin{aligned}
|K_{\phi_{tr}}(S)|^2 &\leq \sum_{j=1}^\infty \lambda_j^{-2}F_{a_j}^2(\phi_{tr})\Norm{\E[q^*(Z,A,X)S(Z,A|W,X)]}_{L_2(Z,A,W,X)}^2 \\ &\lesssim \sum_{j=1}^\infty \lambda_j^{-2}F_{a_j}^2(\phi_{tr})\Norm{ S(Z,A|W,X)}_{L_2(Z,A,W,X)}^2
\end{aligned}
\end{equation*}
Since $\mathcal{R}(T_{tr}')= \{S \in L_2(Z,A,X):\sum_{j=1}^\infty \lambda_j^{-2}\la S, a_j \ra^2 < \infty\}$ by SVD of $T_{tr}'$, $K_{\phi_{tr}}(S)$ is bounded whenever $\phi_{tr} \in \mathcal{R}(T_{tr}')$. Next let, $\phi_{tr}^0 \in \text{cl}(\mathcal{R}(T_{tr}')) - \mathcal{R}(T_{tr}')$. Then $\sum_{j=1}^\infty \lambda_j^{-2}F_{a_j}^2(\phi_{tr}^0) = \infty$. By assumption \ref{regularityq}, for any $r$,
\begin{equation*}
    d_r:=(q^*(Z,A,X)) \Psi_{tr} \sum_{i=1}^r \lambda_i^{-1}F_{a_i}(\psi_{tr})b_i \in L_2(W,Z,A,X).
\end{equation*}
Moreover, $\E[q^*(Z,A,X)) \Psi_{tr} \sum_{i=1}^r \lambda_i^{-1}F_{a_i}(\psi_0)b_i|W,A,X]=0 $ and $\E[q^*(W,a,X)^2 \Psi_{tr} \sum_{i=1}^r \lambda_i^{-1}F_{a_i}(\psi_0)b_i|W,A,X] \in \text{cl}(\mathcal{R}(T_{tr})) $. Thus, $d_r \in \Lambda_2$. Also, we see that $K_{\phi_{tr}}(d_r)=\Norm{d_r}^2$. Then letting $f_r=d_r/\Norm{d_r}$, we see that $f_r$ is a sequence of unit vectors in $\Lambda_2$ such that \begin{equation*}
    \lim_{r\to \infty} K_{\phi_o^t}(f_r)= \left(\sum_{j=1}^\infty \lambda_j^{-2}F_{a_j}^2(\phi_{tr}^0)\right)^{1/2} = \infty,
\end{equation*}
and so $K_{\phi_{tr}^0}(S)$ is unbounded.
\end{proof}
\section{Discussion on Efficiency}
Let $\mathcal{M}$ be the model which places no restriction on the observed data distribution other than existence (but not necessarily uniqueness)
of a bridge function h that solves Equation 1 at all regular laws. We do not compute the semiparametric efficiency bound under this model. Within this model, \cite{semi_proxy} consider the submodel in which the following also hold:
\begin{enumerate}[(a)]
    \item The maps $T_o$ and $T_o'$ are surjective.
    \item Equation 3 holds at the true data generating law.
    \item The $h$ and $q$ that solve 1 and 3 are unique.
\end{enumerate}
They derive that the efficient influence function for the counterfactual mean under the full model $\mathcal{M}$ evaluated at the submodel where a - c hold is $h(W,a,X)-\mu_a+I(A=a)q(Z,a,X)\times(Y-h(W,a,X))$. Thus,
we get the following corollary of Theorem 3.4:
\begin{corollary}
\label{cor:2.6}
Suppose that assumptions 1-5 and equation 15 in the main manuscript hold. The influence function 19 of the de-biased estimator equals the efficient influence function of $\mu_a$ under model $\mathcal{M}$ if a - c also hold at the true data generating law. Thus, the debiased estimator is locally efficient.
\end{corollary}
Importantly, we do not assume a - c for our estimator to be regular and asymptotically linear, nor do we assume a - c hold for all laws in $\mathcal{M}$ for local efficiency, only that they hold at the true data generating law. For more details about local efficiency, we refer the reader to \cite{Bickel1993EfficientAA}. A useful illustration of local efficiency is that of the ordinary least squares estimator. Let $(Y_i,X_i) \sim_{iid} F$ for $F$ unknown. Consider the linear model that only a priori assumes that $E(Y|X)=\beta'X$, but allows the distribution of $Y|X$ to otherwise be unspecified. Then the OLS estimator is regular and asymptotically linear at all laws of the linear model, but also attains the semiparametric efficiency bound for the linear model at the submodel where Var$(Y|X)$ does not depend on $X$, i.e. when the errors are homescedastic. Thus, the least squares estimator is locally efficient. Analogously, our estimator is regular and asymptotically linear at all laws in $\mathcal{M}$, but also attains the semiparametric efficiency bound at the submodel introduced in \cite{semi_proxy}. 
\section{A Larger Class of Functionals}
\label{appendix_class}
Consider the setting from \cite{ghassami}. Let $V$ denote the variables from which independent and identically distributed data is collected. $V_q$ and $V_h$ are (not necessarily disjoint) subsets
of $V$, and $g_1$, $g_2$, $g_3$, and $g_4$ are known measurable functions. Consider estimating a functional that takes the following form:
\begin{equation*}
    \psi := \mathbb{E}[q\left(V_{q}\right) h\left(V_{h}\right) g_{1}(V)+q\left(V_{q}\right) g_{2}(V)+h\left(V_{h}\right) g_{3}(V)+g_{4}(V)],
\end{equation*}
where $h$ and $q$ are nuisance functions and are known to to satisfy 
\begin{equation}
\label{eq:nuisance}
\begin{aligned}
&\mathbb{E}\left[{h}\left(V_{h}\right) g_{1}(V)+g_{2}(V) \mid V_{q}\right]=0, \\
&\mathbb{E}\left[{q}\left(V_{q}\right) g_{1}(V)+g_{3}(V) \mid V_{h}\right]=0,
\end{aligned}
\end{equation}
although the solution to the above equations may not be unique. This includes the functional studied in \cite{nonig}, taking their $\delta$ as $h$ and their $\gamma$ as $q$. Moreover, take $V_h=(X,Z)$, $V_q=(X,Y)$, $g_1(V)=-R$, $g_2(V)=R\tau(X,Y)$, $g_3(V)=(1-R)$, $g_4(V)=R\tau(X,Y)$. In the proximal setting, we consider the counterfactual mean $\mathbb{E}[Y(a)]$. Explicitly, let $V_h = (W,X)$, $V_q = (Z,X)$, and take $h(W,X)$ to be the function that solves $\mathbb{E}[I(A=a)h(W,X)\mid Z,X]=\mathbb{E}[I(A=a)Y\mid Z,X]$ and $q(Z,X)$ to be the function that solves $\mathbb{E}[I(A=a)q(Z,X)-1\mid W,X]=0$. Take $g_1(V)=-I(A=a)$, $g_2(V)=I(A=a)Y$, $g_3(V)=1$, $g_4(V)=0$. One can readily verify that the choices of the $g_i$ and $h$ and $q$ conform to the general class as well as satisfy \ref{eq:nuisance}. The formulation for the proximal setting here does not exactly match the formulation in the main manuscript stylistically. However, the estimation strategy outlined in the main manuscript indeed matches the strategy for the general class outlined below up to some notational differences. Back to the general setting, note that under \ref{eq:nuisance}, 
\begin{equation*}
    \psi = \mathbb{E}[h(V_h)g_3(V)+g_4(V)] =\mathbb{E}[q(V_q)g_2(V)+g_4(V)].
\end{equation*}
Then estimation and inference of $\psi$ can proceed in an analogous manner as in Section 3 of the main manuscript. First, choose to either estimate the solution set for $\mathbb{E}\left[{h}\left(V_{h}\right) g_{1}(V)+g_{2}(V) \mid V_{q}\right]=0$ or \\$\mathbb{E}\left[{q}\left(V_{q}\right) g_{1}(V)+g_{3}(V) \mid V_{h}\right]=0$. For this illustration, suppose we estimate the solution set $\mathcal{H}_0 := \{h \in \mathcal{H}: \mathbb{E}\left[{h}\left(V_{h}\right) g_{1}(V)+g_{2}(V) \mid V_{q}\right]=0\}$. Explicitly, let $\mathcal{H}_n$ be a sieve for $\mathcal{H}$. Let $\widehat{\E}[\cdot \mid V_q]$ be a nonparametric condtional expectation operator and define $e(V_i,h)=\E[h(V_h)g_1(V)-g_2(V)|{V_q}_i]$ and $\wh{e}(V_i,h)=\wh{\mathbb{E}}[h(V_h)g_1(V)+g_2(V)|{V_q}_i]$. Define $C_n(h)=\frac{1}{n}\sum_{i=1}^n \wh{e}^2(V_i,h)$. Let $\widehat{\mathcal{H}}_0=\{ h \in \mathcal{H}_n:C_n(h) \leq c_n\}$ be an estimate for $\mathcal{H}_0$. We can similarly construct an estimate $\widehat{h}_0 \in \widehat{\mathcal{H}}_0$ that converges to a specific element $h_0 \in \mathcal{H}_0$, specifically, $h_0 =\argmin_{h \in \mathcal{H}_0} M(h)$ where $M$ is a convex function, for example $M(h)=\E[h(V_h)^2]$. The sample analog is $M_n(h)=\frac{1}{n}\sum_{i=1}^n h^2({V_h}_{i})$. Then select 
\begin{equation*}
\wh{h}_0\in \argmin_{h \in \wh{\hh}_0}M_n(h).
\end{equation*}
Next, we can construct a plug-in estimator for $\psi$ by taking $\widehat{\psi}=\frac{1}{n}\sum_{i=1}^n g_3(V_i)\widehat{h}_0({V_h}_i)+g_4(V_i)$. Such an estimator will presumably not be root-$n$ consistent. Instead, define the inner product 
\begin{equation*}
    \langle h_1, h_2 \rangle_w=\E[\E[g_1(V)h_1(V_h)|V_q]\E[g_1(V)h_2(V_h)|V_q]].
\end{equation*}
Suppose that the following holds:
\begin{assumption}
There exists a function $h_r \in \mathcal{H}$ such that $\langle h_r, h \rangle_w=\E[h(V_h)g_3(V)]$ for all $h \in \overline{\mathcal{H}}$.
\end{assumption}
This assumption holds if there exists an $h_r$ that satisfies $\E[g_1(V)h_r(V_h)|V_q]=-q(V_q)$ for any $q$ that satisfies the moment equation $\mathbb{E}\left[{q}\left(V_{q}\right) g_{1}(V)+g_{3}(V) \mid V_{h}\right]=0$. To see this, note that
\begin{equation*}
    \begin{aligned}
     \langle h_r, h \rangle_w &= \E[\E[g_1(V)h_r(V_h)|V_q]\E[g_1(V)h(V_h)|V_q]] \\ &= \E[-q(V_q)\E[g_1(V)h(V_h)|V_q]] \\ &= \E[-g_1(V)q(V_q)h(V_h)] \\ &= \E[\E[-g_1(V)q(V_q)h(V_h)|V_h]] \\ &= \E[\E[g_3(V)h(V_h)|V_h]] \\ &= \E[g_3(V)h(V_h)]
    \end{aligned}
\end{equation*}
Under similar regularity conditions as in the main manuscript, it is possible to show that 
\begin{equation*}
    \sqrt{n}(\wh{\psi}-\psi)=\frac{1}{\sqrt{n}}\sum_{i=1}^n g_3(V_i)h_0({V_h}_i)-\E[g_1(V)h_r(V_h) \mid {V_q}_i]\times (h({V_h}_i)g_1(V_i)+g_2(V_i))+g_4(V_i) - \psi -\sqrt{n}r_n(\wh{h}_0)+o_p(1),
\end{equation*}
where $\sqrt{n}r_n(\wh{h}_0)=\frac{1}{n}\sum_{i=1}^n \wh{E}\{g_1(V)h_r(V_h) \mid {V_q}_i\}\wh{e}({V_i},\wh{h}_0)$.
Consider the criterion function
\begin{equation*}
    R(h)=\E[\E[g_1(V)h(V_h) \mid V_q]^2]-2\E[g_3(V)h(V_h)].
\end{equation*}
In a similar vein as in the main manuscript, we can construct a debiased estimator by estimating $r_n(\wh{h}_0)$. First, let \begin{equation*}
    R_n(h)=\frac{1}{n}\sum_{i=1}^n\wh{\E}\{g_1(V)h(V_h)|V_{q_i}\}^2-\frac{2}{n}\sum_{i=1}^n g_3(V_i)h(V_{h_i}) \ , \ h \in \hh
\end{equation*}
be the sample version of the criterion function $R(h)$.
Observe that since $\la h_r, h \ra_w=\E\{g_3(V)h(V_h)\}$, we have that $R(h)=\Norm{h-h_r}^2_w-\Norm{h_r}_w^2$. It follows that $h_r$ is the unique minimizer of the mapping $h \to R(h)$. Then  we can estimate the term $h_r$ by 
\begin{equation}
    \wh{h}_r \in \argmin_{h \in \hh_n } R_n(h).
\end{equation}
With this estimate, we can construct the following estimator for $r_n(\wh{h}_0)$
as 
\begin{equation}
    \wh{r}_n(\wh{h}_0)=\frac{1}{n}\sum_{i=1}^n  \wh{\E}\{g_1(V)\wh{h}_r(V_h)|V_{q_i}\}\wh{e}(V_i,\wh{h}_0).
\end{equation}
The debiased estimator
$\wh{\psi}_{db}=\wh{\psi}+ \wh{r}_n(\wh{h}_0)$  can be shown to have influence function $g_3(V)h_0({V_h})-\E[g_1(V)h_r(V_h) \mid {V_q}]\times (h({V_h})g_1(V)+g_2(V))+g_4(V) - \psi$ under similar regularity conditions as in the main manuscript. If $\E[g_1(V)h_r(V_h) \mid {V_q}] = -q(V_q)$, this equals $$g_3(V)h_0({V_h})+q(V_q) h_0({V_h})g_1(V)+g_2(V)q(V_q)+g_4(V) - \psi,$$
which corresponds to the influence function for functionals studied in \cite{ghassami} if $h$ and $q$ are unique.

\bibliographystyle{abbrvnat}
%\bibliography{template_short}
\bibliography{zupplement.bib}